\magnification \magstep1
\raggedbottom
\openup 2\jot
\voffset6truemm

\centerline {\bf REVISITED VERSION OF WEYL'S LIMIT POINT-LIMIT CIRCLE CRITERION}
\centerline {\bf FOR ESSENTIAL SELF-ADJOINTNESS}

\vskip 1cm
\leftline {\it Vito Flavio Bellino}
\vskip 0.3cm
\noindent {Universit\`a di Napoli Federico II, Dipartimento di Fisica ``Ettore Pancini'',  
Complesso Universitario di Monte Sant'Angelo, Via Cintia 21 - 80126 Napoli, Italy}
\vskip 1cm

\leftline {\it Giampiero Esposito}
\vskip 0.3cm
\noindent {INFN Sezione di Napoli, Complesso Universitario di Monte S. Angelo, 
Via Cintia Edificio 6, 80126 Napoli, Italy}
\vskip 1cm

{\bf Abstract.}
The principal aim of this paper is to present a new proof of Weyl's criterion 
in which it is shown that the natural framework for the associated 
Sturm-Liouville operators is $W^{2,1}\cap L^{2}$ -i.e.- the intersection of a particular 
Sobolev space and of the $L^{2}$ space. Indeed, we will deal with the special case of the radial operator 
$\left(- {d^{2} \over dx^{2}} + q(x)\right)$ on a real line segment 
(either bounded or unbounded) that often occurs 
in the study of quantum systems in central potentials. 
We also derive from first principles the functional behaviour of the coefficients 
for a general second-order Sturm-Liouville operator by using some extensions 
of a milestone Carath\'eodory existence theorem.
\vskip 100cm
\leftline {\bf 1. Introduction}
\vskip 0.3cm
\noindent
Limit-Point Limit-Circle theory was first developed by the young Herman Weyl in the early 1900's in one of his first 
articles [1]. Since then, such methods (hereafter denoted by LP and LC, respectively) 
have become increasingly important thanks to their accurate predictions on the form of the
potential in the applications, which can easily supply foundamental information about the solution of a great 
variety of singular second-order Sturm-Liouville problems. 
In the modern literature, the work in [2,3] provides an enlightening introduction to the
link between these singular second-order problems and functional analysis, as well as to the
applications to ordinary quantum mechanics.

The world of atomic physics offers indeed a wide range of applications of ordinary quantum
mechanics. This is not an exact theory, because relativity would make it
necessary to use the spectral theory of pseudo-differential operators in order to
develop the quantum theory of bound states [4]. Thus, one still resorts with profit to
ordinary quantum mechanics, from which one can learn valuable lessons. For example, if 
a physical system ruled by a central potential $V(r)$ is considered in ${\bf R}^{n}$
(the choice $n=3$ is frequent but not mandatory), one finds an effective potential
$$
V_{\rm eff}(r)=V(r)+{\rho_{nl} \over r^{2}},
\eqno (1.1)
$$
where, on denoting by $l$ the orbital angular momentum quantum number, one finds [5]
$$
\rho_{nl}={(n-1)(n-3)\over 4}+l(l+n-2)
=\left(l+{(n-2)\over 2}\right)^{2}-{1 \over 4}.
\eqno (1.2)
$$
As one learns from [2,3], the LP condition at the origin is achieved if
$$
V(r)+{\rho_{nl}\over r^{2}} \geq {3 \over 4r^{2}}
\; \; {\rm as} \; \; r \rightarrow 0^{+}.
\eqno (1.3)
$$
In the particular case of a free particle, $V(r)=0$ and (6.3) leads to
$$
\left(l+{(n-2)\over 2}\right)^{2}-{1 \over 4} \geq {3 \over 4} 
\Longrightarrow l+{n \over 2} \geq 2.
\eqno (1.4)
$$
Interestingly, this condition is violated just once, i.e. by $s$-wave stationary
states (for which $l=0$) in $3$ dimensions. The same holds if $V(r)$ is a Coulomb-type potential,
because then the centrifugal term on the left-hand side of (1.3) dominates on the
Coulomb term as $r$ approaches $0$.

Section 2 describes what is known from a Carath\'eodory theorem on ordinary differential
equations; section 3 studies the Sobolev functional space for solutions of our
singular second-order problems; Weyl's LP-LC criterion is studied with extensive and original use of
Sobolev spaces in sections 4 and 5. 
Explicit examples of self-adjoint extensions are analyzed in section 6.
Concluding remarks are made in section 7, while relevant
details are given in the appendix. Throughout our paper, the reader is assumed to have 
some background on the LP-LC theory [6,7] and on operator theory [2,3].
\vskip 0.3cm
\leftline {\bf 2. Extended Carath\'eodory's existence Theorem}
\vskip 0.3cm
\noindent
First, we want to understand the functional behaviour of the coefficients 
and all possible solutions of the general 
Sturm-Liouville eigenvalue equation
$$
-{d\over dx}\biggl(p(x){dy\over dx}\biggr)+q(x)y=ly\hskip1cm l\in {\bf C}.
\eqno (2.1)
$$
In the LP-LC literature, whenever one deals with equation (2.1), it is necessary 
to specify the functional space to which 
the coefficients $p(x),\;p'(x)$ and $q(x)$ belong, in order to develop the theory and reach the desired results.

\noindent For example, in [6,7,8] the coefficients $p,p'$ and $q$ belong 
to the set of real-valued continuous functions and hence 
the solution $y$ must be globally of class $C^{2}$ on the interval 
$I\subseteq \bf R$ of interest. On the other 
hand, in [9,10], weaker conditions on $p^{-1}, q$ are given -i.e.- 
they are $L^1_{loc}$ while the solution $y$ and its 
derivative $y^{'}$ are absolutely continuous ($AC_{loc}$) on the interval $I$ of interest.
\vskip0.2cm
In order to clarify the hypotesis made on such functions and on the solutions, 
we will make use of some extended Carath\'eodory's existence theorems:
\vskip0.5cm
\noindent
{\bf Theorem 2.1.}\hskip0.5cm {\it Let $I\subset\bf R$ be a closed interval and 
let $f\bigl(x,u(x)\bigr)$ be G-regular on $I$ (see appendix). 
Then there exists at least one absolutely continuous function $u$ such that 
$$
u(x)=\int^{x}_{x_0}f\bigl(s,u(s)\bigr)ds\hskip1cm x\in I
\eqno (2.2)
$$
where $x_0$ is the average point of $I$.}
\vskip0.5cm
\noindent We note that if $f$ obeys the above theorem, there exists 
at least one absolutely continuous function $u$ that 
satisfies the equation $u'=f\bigl(x,u(x)\bigr)$ almost everywhere.
\vskip0.5cm
\noindent
{\bf Theorem 2.2.}\hskip0.5cm {\it Let $I=[a,b]\subset\bf R$ be a compact interval 
and let $f\bigl(x,y\bigr): I\times V\rightarrow V$ 
satisfy the following hypothesis (where $V$ is a generic $n$-dimensional space):\par
\vskip0.2cm
1) \hskip0.5cm For every $y\in V$, $f(x,y)$ is measurable on $I$ and it is continuous in $V$.\par
2) \hskip0.5cm There exists a real-valued non-negative function 
$M(x)\in L^{1}(I)$ such that 
$$
|f(x,y)|\leq M(x).
$$
\vskip0.2cm
\noindent Then there exists an absolutely continuous function $u(x)$ 
such that $u'(x)=f\bigl(x,u(x)\bigr)$ almost everywhere on $I$.}
\vskip0.5cm
Theorem 2.1 is discussed and proved in [8] while Theorem 2.2 in [11].

\noindent In [11] it is shown that the requirement 2) of Theorem 2.2 can be replaced by the following:
\vskip0.2cm
{\it2')} \hskip0.5cm {\it For every $y\in C(I)$, $f\bigl(x,y(x)\bigr)$ is 
summable in $I$, and upon taking $y\in C(I)$, 
\par the functions $\displaystyle\int_a^{x}f\bigl(t,y(t)\bigr)dt$ describe 
an {\it absolutely equicontinuous} family on $I$}
\vskip0.2cm
\noindent obtaining a more general existence theorem.

\noindent Here we want to show that, if the assumptions of Theorem 2.1 are verified 
and we also take hypothesis {\it 2')} instead of {\it 2)}, then 
Theorem 2.2 must also be true. From this the former will be a restricted case of the latter. 

\noindent As already mentioned, for the definition of {\it G-regularity} 
we remind to the Appendix at the end of this paper. Here we will only give 
the main condition that ensures the occurrence of this property. 
For this purpose we need some further notions: 
\vskip0.5cm
\noindent
{\bf Definition 1.}\hskip0.5cm Let $I\subset {\bf R}$ and let 
$h^{1},...,h^{m},k^{1},...,k^{m}\in L^{1}(I,{\bf R})$. We define 
the subsequent convex subset $G(h,k)$ of $L^{1}(I,{\bf R}^{m})$ by
$$ 
G(h,k)=\{ g\in L^{1}(I,{\bf R}^{m}),\; h^{j}(x)\leq g^{j}(x)
\leq k^{j}(x),\; x\in I,\;1\leq j\leq m \}.
\eqno (2.3)
$$
{\bf Theorem 2.3.}\hskip0.5cm {\it If the function $f:(x,y)\in I\times{\bf R}^{m}
\rightarrow f(x,y)\in {\bf R}^{m}$ satisfies
$$
|f(x,y)|\leq M(x)(1+|y|),\hskip1cm(x,y)\in I\times{\bf R}^{m}
\eqno (2.4)
$$
for some $M\in L^{1}(I,{\bf R})$ and it is measurable in the $x$ variable 
for any fixed $y$ and it is also continuous in 
the $y$ variable, then there exists a $G(h,k)$ such that $f$ is G-regular on $I$.
}

\vskip0.5cm
Of course, form Theorem 2.3 the assumptions of Theorem 2.1 are satisfied 
and hence (2.2) is absolutely continuous.
\vskip0.5cm
\noindent
{\bf Definition 2.}\hskip0.5cm Let $G=G(h,k)$ be given by (2.3) and $x_0$ be 
the middle point of the interval $I$. 
Let $f:I\times{\bf R}^{m}\longrightarrow{\bf R}^{m}$ be such that
$$
f\biggl(x,\int_{x_0}^{x}g(s)ds\biggr)\in G\hskip1cm for\;all\;g\in G
\eqno (2.5)
$$
Then $f$ is said to be {\it G-integrable} on $I$.
\vskip0.5cm
As we can see from the definition of G-regularity in our appendix, the 
G-integrability is necessary for the G-regularity. 
By using some content in [12] we can easily see that the G-integrability 
implies the absolute continuity of 
$\int_{{x_0}}^{x}f\bigl(t,y(t)\bigr)dt$ in the $x$ variable where $y$ is taken 
to be absolutely continuous as in (2.4).

\noindent From this, one finds that the G-regularity makes (2.2) absolutely 
continuous whenever a particular absolutely continuous function $u(x)$ is chosen, 
and hence $\int_{{x_0}}^{x}f\bigl(s,u(s)\bigr)ds$ is a family of absolutely continuous 
functions if we let the $u$ variable run over a particular set of continuous functions.

\noindent Since under our hypothesis $f$ is taken G-integrable, it is also 
bounded from the Definition 2 and this suggests us 
that $ \int_{x_{0}}^{x}f\bigl(s,u(s)\bigr)ds$ is an equi-absolutely continuous 
family of functions. From this we have already 
proved that Theorem 2.1 is a special case of Theorem 2.2 when the {\it 2)} 
hypothesis is replaced with {\it 2')}.
\vskip 0.3cm
\leftline {\bf 3. Sobolev functional space for solutions}
\vskip 0.3cm
\noindent
Now, by expressing (2.1) in the subsequent form of first-order differential system:
$$ 
\cases{ {dy\over dx}=v \cr\cr 
{dv\over dx}=-{p^{'}(x)\over p(x)}v+{q(x)\over p(x)}y-{l\over p(x)}y \cr}
\eqno (3.1)
$$
it is easy to see, by applying Theorem (2.2) with the {\it 2')} hypothesis, 
that the required summability in the $x$ variable 
forces the coefficients $p(x)^{-1},\,p'(x)$ and $q(x)$ to belong
to $L^{1}(I)$ while the solution $y$ to (2.1) and its 
derivative $y'$ are absolutely continuous functions.

Now we will face the fact that an absolutely continuous function must belong to 
a Sobolev space -i.e.- $W^{1,1}$ defined, for example, in [13]:
\vskip0.5cm
\noindent
{\bf Definition 3.}
$$
W^{1,1}([a,b]) \equiv
\biggl\{ u\in L^{1}(a,b):\exists g\in L^{1}(a,b): 
\int_{a}^{b} u \phi'
=-\int_{a}^{b} g\phi \biggr\},
\eqno (3.2)
$$
for all $\phi \in C_{0}^{1}(a,b)$. We also recall the following
\vskip0.5cm
\noindent
{\bf Definition 4.}\hskip0.5cm Let $f:{\bf R}\longrightarrow{\bf R}$ be a function in 
$[a,b]$ such that its derivative is defined almost everywhere and
$$
\displaystyle\int_a^b{df\over dx}dx=f(b)-f(a),
\eqno (3.3)
$$
then $f$ is said to be absolutely continuous.
\vskip0.5cm

\noindent It is well known that the weaker classical hypothesis that makes it possible
to perform an integration by parts (Lebesgue fondamental integral theorem) such as
$$
\int_a^b f {d\phi\over dx}=-\int_a^b{df\over dx}\phi\hskip1cm \phi\in C^{1}_0(a,b)
\eqno (3.4)
$$
is the {\it absolute continuity} of the function $f$.

\noindent Now, from Definition 5, it follows that such functions must be continuous and 
have got bounded variation in the compact interval $[a,b]$. 
This simpy implies that every such function belongs to the $L^1(a,b)$ space. 
Furthermore, from the bounded variation behaviour 
of $f$, the integral (3.5) must be finite for every compact interval $[a,b]$, 
thus $f'$ must belong to $L^{1}(a,b)$.

\noindent Taking into account the Definition 3, jointly with the properties 
obtained above, we see that every absolutely 
continuous function belongs to the $W^{1,1}([a,b])$ Sobolev space.
\vskip 0.3cm
\leftline {\bf 4. Weyl's LP-LC criterion}
\vskip 0.3cm
\noindent
Let us consider the following special case of Sturm-Liouville equation on 
$(a,b)$ taking $p=1$ in (2.1)
$$
-{d^{2}y\over dx^2}+q(x)y=ly\hskip1cm l\in {\bf C} .
\eqno (4.1)
$$
This is an eigenvalue equation whose differential operator is
$$
{\cal L}=-{d^{2}\over dx^{2}}+q(x)
\eqno (4.2)
$$
defined on $L^{2}(a,b)$. The aim of the following {\it Weyl's Criterion} 
is to provide the condition on the operator (4.2) in order 
to ensure its self-adjointness in terms of the LP-LC property. 
In this way, such a theorem provides a magnificent link 
between operator theory on Hilbert spaces and LP-LC theory [2,3]:
\vskip0.5cm
\noindent
{\bf Theorem 4.1.}\hskip0.5cm {\it Let $q(x)\in L^{2}_{\rm loc}$ in $I=(a,b)$ and let 
${\cal L}={\displaystyle -{d^{2}\over dx^{2}}}+q(x)$ 
with domain $D({\cal L})= C_{0}^{\infty}(a,b)$. Then the closure 
${\bar{\cal L}}$ has got deficiency indices:
\vskip0.3cm

(i)\hskip0.5cm $n_{+}(\bar{\cal L})=n_{-}(\bar{\cal L})=2$ if $\cal L$ is in LC at both ends of the interval;

(ii)\hskip0.4cm $n_{+}(\bar{\cal L})=n_{-}(\bar{\cal L})=1$ if $\cal L$ is in LC at one end and LP at the other;

(iii)\hskip0.3cm $n_{+}(\bar{\cal L})=n_{-}(\bar{\cal L})=0$ if $\cal L$ is in LP at both ends of the interval.
\vskip0.3cm
Therefore, $\cal L$ is essentially Self-Adjoint if and only if it is LP at both end-points of the interval.}

\vskip0.5cm
From the extended version of Carath\'eodory existence theorem we know that, 
if a solution to equation (4.1) exists in a 
compact $I$, then it must be absolutely continuous together with its first 
derivative and thus it belongs to $W^{1,1}(I)$.

\noindent We can introduce the following Sobolev space that will be the 
basic living place for our solutions:
\vskip 0.5cm
\noindent
{\bf Definition 5.}
$$
W^{2,1}(I)=\{ u\in W^{1,1}(I)\;|\;u' \in W^{1,1}(I)\}.
\eqno (4.3)
$$
\vskip0.5cm
Of course, we are only interested in functions which are absolutely continuous 
with their first derivative.

\noindent We note that the operator (4.2) must act on a Hilbert space 
-i.e.- $L^{2}(I)$ and therefore, from (4.1), 
the function ${d^{2}y \over dx^{2}}\in L^{2}(I)$. 

\noindent The subsequent theorem [11] shows that the hypothesis on the second 
derivative of our solutions to belong to $L^{2}$ 
is sufficient to guarantee us the {\it local absolute continuity} of 
the solutions and their first derivative:
\vskip0.5cm
\noindent
{\bf Theorem 4.2.}\hskip0.5cm{\it  Let $g\in L_{\rm loc}^{1}(I)$ and take for some 
$y_{0}\in I$ the following expression:
$$
v(x)=\int_{y_{0}}^{x}g(t)dt\hskip1cm x\in I.
\eqno (4.4)
$$
Then, $v(x)$ is continuous in $I$ and
$$
\int_{I}v\phi^{'}=-\int_{I}g\phi\hskip1cm \forall\phi\in C^{1}_{0}(I).
\eqno (4.5)
$$
}
\vskip0.5cm
By applying recursively the above theorem one finds that, 
under the hypothesis of square summability of its second derivative, 
$y$ and its first derivative are {\it locally absolutelly continuous} on $I$ 
whatever $I$ is. It is also clear that, in the case of a compact real interval, 
$y$ and $y'$ are absolutely continuous functions of $W^{1,1}(I)$ and hence 
$y$ belongs to the (4.3) set.

\noindent Now, if $I$ is not bounded or half-bounded, we are dealing 
with functions belonging to $W^{2,1}_{\rm loc}(I)$ 
and taking the square summability required for operator (4.2), the basic 
functional space to which our solution belongs is $W^{2,1}_{\rm loc}(I)\cap L^{2}(I)$.
\vskip0.5cm
Now we can summarise our results in the following theorem:
\vskip0.5cm
\noindent
{\bf Theorem 4.3.}\hskip0.5cm Whatever the $I$ interval is, every solution to equation 
(4.1) belongs to the space $W^{2,1}_{\rm loc}(I)$.
\vskip0.5cm
Of course, if $I$ is compact then $W_{\rm loc}^{2,1}(I)\equiv W^{2,1}(I)$ while if the square summability is required, 
then $y''$ must belong to $L^{2}(I)$ and thus we obtain the {\it local absolute continuity} of $y$ and $y^{'}$ and 
the Carath\'eodory's existence theorem is fullfilled.
\vskip 0.3cm
\leftline {\bf 5. Proof of Weyl's criterion}
\vskip 0.3cm
\noindent
We can now proceed with the proof of Theorem 4.1 by following the logical steps 
that can be found in [3]. Our method will make use 
of Theorem 4.2 jointly with all the information obtained in the previous section.
\vskip0.5cm
\noindent
Proof of statement (i):
\vskip0.5cm
If the operator $\cal L$ is LC at both ends of the interval $I$, then every solution to the equation ${\cal L}y=ly$, 
$\forall l\in {\bf C}$ for which $\Im l\not=0$, belongs to $L^{2}(I)$. 
This means that there exist two linearly independent solutions 
to each of the equations ${\cal L}y=iy$ and ${\cal L}y=-iy$, and therefore the deficiency indices 
are $n_{+}(\bar{\cal L})=n_{-}(\bar{\cal L})=2$.
\vskip0.5cm
\noindent
Proof of statement (ii):
\vskip0.5cm
Suppose that ${\cal L}$ is LP at $a$ and LC at $b$. 

\noindent Let us consider a restriction ${\cal L}_{0}$ of the operator 
$\cal L$ acting on the subsequent linear domain:
$$
D({\cal L}_{0})=\{\phi\in W_{\rm loc}^{2,1}([c,d]):\phi(c)=\phi(d)
=\phi'(c)=0,\;\phi'' \in L^{2}(c,d)\}
\eqno (5.1)
$$
where $[c,d]\subset I$. From Theorem 4.2 and the comment below, we have that $D({\cal L}_{0})
\subset \bigl(W^{2,1}_{\rm loc}([c,d])\cap L^{2}(c,d)\bigr)$, 
hence $D({\cal L}_{0})$ is a Hilbert sub-space of $L^{2}(c,d)$.
It is easily seen that operator ${\cal L}_{0}$ is symmetric and that the domain of its adjoint is
$$
D({\cal L}_{0}^{*})=\{\psi\in W_{loc}^{2,1}([c,d])\;|\;\psi(d)=0,\;\psi'' \in L^{2}([c,d])\},
\eqno (5.2)
$$
because the equations 
$$
{\cal L}_{0}\phi=\pm i\phi
\eqno (5.3)
$$
have at most two linearly independent solutions in $L^{2}(c,d)$ and hence 
$n_{+}({\cal L}_{0})\leq 2$ and $n_{-}({\cal L}_{0})\leq 2$. 
We must rule out the case $n_{+}({\cal L}_{0})=n_{-}({\cal L}_{0})=0$ because 
it is the self-adjoint one and this is 
not the case because $D({\cal L}_{0})\subset D({\cal L}^{*}_{0})$.

\noindent Now we can show that each of equations (5.3) has only one solution in $L^{2}(c,d)$.

\noindent It is indeed well known that the adjoint domain 
for a linear operator on a Hilbert space admits the following decomposition:
$$
D({\cal A}^{*})= D({\cal A})\oplus K_{+}({\cal A})\oplus K_{-}({\cal A})
\eqno (5.4)
$$
where $K_+({\cal A})$ and $K_-({\cal A})$ are the deficiency spaces of the 
operator under consideration. Let us define the operator 
${\cal P}^{2}=- {d^{2}\over dx^{2}}$ on the domain (5.1) 
and let us denote it by ${\cal P}^{2}_{0}$. Of course 
${\cal P}_{0}^{2}$ is symmetric and its adjoint has domain (5.2), thus 
we can certainly say that $K_{+}({\cal L}_{0})
\oplus K_{-}({\cal L}_{0})=K_{+}({\cal P}^{2}_{0})\oplus K_{-}({\cal P}^{2}_{0})$. 
By solving the equations
$$
{\cal P}^{2}\phi=\pm i\phi
\eqno (5.5)
$$
with the condition imposed by (5.2), we obtain two one-dimensional deficiency spaces of the form  
$$
\eqalignno{ 
K_{+}({\cal P}^{2}_{0}) & = \biggl\{ \xi \in L^{2}(c,d)\;|\; \xi(x)=\lambda\biggl(\zeta_{1}(x)
-{\displaystyle{\zeta_{2}(x)\over e^{\sqrt2(i+1)d}}}\biggr)\;|\;\lambda\in{\bf C}\biggr\}.\cr
& K_{-}({\cal P}^{2}_{0})= \biggl\{ \eta \in L^{2}(c,d)\;|\; \eta(x)
=\lambda\biggl(\rho_{1}(x)-{\displaystyle{\rho_{2}(x)
\over e^{\sqrt2(i-1)d}}}\biggr)\;|\;\lambda\in{\bf C}\biggr\}.
&(5.6)\cr}
$$
where $\zeta_{1}$, $\zeta_{2}$, $\rho_{1}$ and $\rho_{2}$ are locally square integrable on the real line.
\noindent Equations (5.6) show that $K_{+}({\cal P}^{2}_{0})\oplus K_{-}({\cal P}^{2}_{0})$ is a two-dimensional 
linear space and hence the same holds for $K_{+}({\cal L}_{0})\oplus K_{-}({\cal L}_{0})$.
\noindent Now, taking into account the fact that [6,7] if, for some complex $l_{0}\in {\bf C}$ all solutions to 
${\cal L}_{0}y=l_{0}y$ are square integrable, than for every complex 
$l\in{\bf C}$ every solution to ${\cal L}_{0}y=ly$ 
is square integrable as well, we must rule out the cases ${\rm dim}(K_+({\cal L}_{0}))=2$, 
${\rm dim}(K_{-}({\cal L}_{0}))=0$ and ${\rm dim}(K_+({\cal L}_{0}))=0$, 
${\rm dim}(K_{-}({\cal L}_{0}))=2$. From this, the only case left is ${\rm dim}(K_{+}({\cal L}_{0}))
={\rm dim}(K_{-}({\cal L}_{0}))=1$ and therefore $n_{+}({\cal L}_{0})=n_{-}({\cal L}_{0})=1$.
\noindent This also means that there exists a non-vanishing function $\tilde u$ 
that does not belong to ${\rm Ran}({\cal L}_{0}-i{\cal I})$.
At this stage, let $I_{-}=(a,d]$ and let ${\cal L}_{1}$ be a second restriction of ${\cal L}$ defined on
$$
\eqalignno{ D({\cal L}_{1}) & =\{\phi\in W_{\rm loc}^{2,1}(I_{-})\cap L^{2}(I_{-})
\;|\;\phi(x)=\phi(d)=0,\;x\in(a,a+\epsilon),\cr
& \hskip3.8cm\epsilon\in(a,d),\;\phi'' \in L^{2}(c,d)\}.
&(5.7)\cr}
$$
We note that in the case in which the end-point $a$ is at finite distance from the origin, 
the basic space in (5.7) can always be taken to be 
$W^{2,1}(I_{-})$ instead of $W^{2,1}(I_{-})\cap L^{2}(I_{-})$.

\noindent All our reasoning on ${\cal L}_{0}$ can be repeated on ${\cal L}_{1}$, leading us to the same conclusions: 
${\cal L}_{1}$ is symmetric on its domain and has got the same deficiency indices 
of ${\cal L}_{0}$. By using some arguments that can be found in [2], we can state 
that there exists at most one self-adjoint extension of ${\cal L}_{1}$ by virtue of the 
equality of its deficiency indices. Let us denote by ${\cal L}_{2}$ such a self-adjoint extension.
\noindent It is clear that, if the domain is taken to be
$$
D({\cal L}_{2})=\{\phi\in W_{\rm loc}^{2,1}(I_{-})\cap L^{2}(I_{-});|\;\phi(a)=\phi(d)=0,\;\phi'' \in L^{2}(I_{-})\}
\eqno (5.8)
$$
we are dealing with a self-adjoint extension of ${\cal L}_{1}$ and we also note that 
$D({\cal L}_2)\subset D(\bar{\cal L}_{1})$.

\noindent Take now a function $\chi\in D({\cal L}_{2})$ for which ${\cal L}_{2}\chi-i{\cal I}\chi=u$ where $u$ is chosen 
in such a way that its restriction $\tilde u$ to the interval $[c,d]$ is not in ${\rm Ran}({\cal L}_{0}-i{\cal I})$. 
\noindent We easily see that $\chi$ cannot be equal to zero on $(a,c)$. To see this we have to take into account that 
$\chi\in C^{1}$ on $(a,d]$ (we refer to the previous section); 
if this were possible, then the restriction $\tilde \chi$ 
of $\chi$ to the interval $[c,d]$ would belong to $D({\cal L}_{0})$ 
because we would have $\chi(c)=\chi(d)=\chi'(a)=0$ 
and one would find that ${\cal L}_{2}{\tilde \chi}-i{\cal I}{\tilde \chi}={\tilde u}$. But this contradicts our previous 
hypothesis, thus $\chi$ cannot be equal to zero on $(a,c)$.

\noindent Last, since $\chi$ is $L^{2}$ near the $a$ end-point in LP and the 
operator ${\cal L}$ is LC near the $b$ end-point, 
the continuous extension $\hat \chi$ of $\chi$ that solves the equation 
${\cal L}y= iy$ over the whole $(a,b)$ is the only $L^{2}(a,b)$ 
solution. The same holds for the equation ${\cal L}y= -iy$ and hence we have 
$n_{+}(\bar{\cal L})=n_{-}(\bar{\cal L})=1$.
\vskip 0.3cm
\noindent
Proof of statement (iii):
\vskip 0.3cm
Suppose that ${\cal L}$ is LP at both end-points $a$ and $b$.
From the (ii) statement we know that there exists only one square-integrable 
function near $a$ and one near $b$ but 
we do not know whether they can be related in some way. 
Instead of showing that this is not the case, we will show that the 
deficiency indices of $\bar{\cal L}$ are $n_{+}(\bar{\cal L})=n_{-}(\bar{\cal L})=0$.

\noindent In order to do this, we first need some arguments on the Wronskian function
$$
{\cal W}(x;\phi,\psi)=(\phi\psi'-\phi' \psi)(x)\hskip1cm x\in I,\;\;\;\phi,\psi\in D({\cal W})
\eqno (5.9)
$$
where $D({\cal W})$ is defined according to
$$
D({\cal W})=\{\varphi \in W_{\rm loc}^{2,1}(I)\;|\;\varphi'' \in L^{1}_{\rm loc}(I)\},
\eqno (5.10)
$$
and on the {\it regular points} of the operator $\cal L$.
\noindent We say that the point $a$ is a {\it regular point} (the same for $b$) for $\cal L$ 
if and only if it is finite and the subsequent condition holds:
$$
\int _{a}^{d}|q(x)|^{2}<\infty\hskip1cm\forall d\in I.
\eqno (5.11)
$$
\vskip0.5cm
\noindent First we want to show that: \par
\vskip0.1cm
(a)\hskip0.5cm ${\cal W}(x;\phi,\psi)$ is a locally absolutely continuous function in the $x$ variable.\par
\vskip0.1cm
(b)\hskip0.5cm If $\phi,\psi\in D({\cal L}^{*})$ then there exist the limits 
$\displaystyle\lim_{x\rightarrow a}{\cal W}(x;\phi,\psi)$ 
and $\displaystyle\lim_{x\rightarrow b}{\cal W}(x;\phi,\psi)$ and hence
$$
{\cal W}(b;\bar\phi,\psi)-{\cal W}(a;\bar \phi,\psi)=<\phi,{\cal L}^{*}\psi> - <{\cal L}^{*}\phi,\psi>
\eqno (5.12)
$$
\par
(c)\hskip0.5cm The operator $\cal L$ is in LC at its regular point, 
and if $\phi\in D({\cal L}^{*})$ then the limits\par
$\hskip 0.9cm \displaystyle \lim_{x\rightarrow a}\phi(x)$ and 
$\displaystyle \lim_{x\rightarrow a}\phi'(x)$ 
exist and are zero if $\phi\in D(\bar{\cal L})$.
\vskip 0.3cm
\noindent 
In order to prove the (a) property, we only 
have to take into account Theorem 4.2 that ensures the local absolute continuity 
of the functions in $D({\cal W})$ and of their first derivative. From the 
definition (5.9) it is easily seen that the product and summation of 
locally absolutely continuous functions lead to the local abolute continuity of the Wronskian.  

\noindent For point (b) we can use the fact that the function $q(x)\phi\psi\in L^{2}_{\rm loc}(I)$ because 
$\phi,\psi\in D({\cal W})$, while $q(x)\in L^{2}_{\rm loc}(I)$ from the hypothesis.
Now, taking into account Theorem 4.3 and the fact that the functions of $D({\cal L}^{*})$ need the square integrability 
of their second derivatives, from the H{\"o}lder inequality, such derivatives are also locally integrable and our 
starting functions belong to $D({\cal W})$. From this we have that 
$D({\cal L}^{*})\subseteq (D({\cal W})\cap L^{2}(I))$.

\noindent Using the local absolute continuity of ${\cal W}$, we have for all $[c,d]\subset I$ that
$$
{\cal W}(d;\bar\phi,\psi)-{\cal W}(c;\bar\phi,\psi)
=\int_{c}^{d}({\bar \phi}\psi''-{\bar \phi}''\psi),
\eqno (5.13)
$$
and on adding and subtracting the $\displaystyle \int_{c}^{d}q(x)\phi\psi$ term in (5.13), 
under the hypothesis of $\phi,\psi\in D({\cal L}^{*})$, we have
$$
{\cal W}(d;\bar\phi,\psi)-{\cal W}(c;\bar\phi,\psi)
=\int_{c}^{d}(\overline\phi{{\cal L}^{*}\psi}-\overline{{\cal L}^{*}\phi}\psi).
\eqno (5.14)
$$
Since the functions in $D({\cal L}^{*})$ are locally absolutely continuous and globally square integrable 
with their second derivative on $I$, in (5.14) the limits 
$$
\displaystyle\lim_{c\rightarrow a}{\cal W}(x;\phi,\psi)
\; {\rm and} \; 
\displaystyle\lim_{d\rightarrow b}{\cal W}(x;\phi,\psi)
$$ 
exist and equation (5.12) holds. This completely proves the (b) Wronskian statement.

\noindent As far as the property (c) is concerned, under
the hypothesis of regularity for point $a$, if $\phi\in D({\cal L}^{*})$, 
then it is in $D({\cal W})$ and therefore $\phi$ and $\phi'$ are 
locally absolutely continuous. This ensures us that 
the limits $\displaystyle \lim_{x\rightarrow a}\phi(x)$ and 
$\displaystyle \lim_{x\rightarrow a}\phi'(x)$ must exist. 

\noindent Now, since $D({\cal L})\equiv C_{0}^{\infty}(I)$ and the map 
$\phi\rightarrow(\phi(a),\phi'(a))$ is continuous 
in the norm $||\phi||+||{\cal L}^{*}\phi||$, then 
$\displaystyle \lim_{x\rightarrow a}\phi(x)=\displaystyle \lim_{x\rightarrow a}
\phi'(x)=0$ for all $\phi\in D(\bar {\cal L})$. Since $\phi$ is locally absolutely continuous, 
it is bounded near $a$ and thus it belongs to $L^{2}$. This shows the LC case.
\vskip 0.3cm
\noindent We can now proceed with the proof of the (iii) statement, 
in which we will make use of the results obtained above.

\noindent Suppose that the end-point $a$ is regular while $b$ is LP. 
It is easily seen that $\cal L$ has got self-adjoint extensions 
because it is symmetric on $C^{\infty}_{0}(I)$ and it has deficiency indices $n_{+}({\cal L})=n_{-}( {\cal L})=1$ 
like the operator ${\cal L}_{1}$ defined in the proof of the (ii) statement.  
Among all conceivable self-adjoint extensions, we want to choose that one 
for which there exists some $(\alpha,\beta)\in {\bf R}^{2}\backslash(0,0)$ such that 
$$ 
\alpha\phi(a)+\beta\phi'(a)=0 
\eqno (5.15)
$$
\noindent and call it $\hat{\cal L}$. In order to do this we will use a 
Theorem in [14], known under the name of {\it von Neumann's 
extension Theorem}\footnote {*}{Let $A$ be a closed Hermitian operator with domain
$D(A)$ dense in a Hilbert space ${\cal H}$, and let us define the spaces [2,3,14]
$$
{\cal K}_{\pm}={\rm Ker}(A^{*} \mp iI),
$$
with dimension denoted by $d_{\pm}$. For any closed symmetric operator $B$, we denote
by $U_{B}$ its Cayley transform
$$
U_{B} \equiv (B-iI)(B+iI)^{-1},
$$
extended to ${\cal H}$ by setting it to $0$ on
${\rm Ran}(B+iI)^{\perp}$. The von Neumann Extension Theorem can be stated
as follows [2,3,14]:
\vskip 0.3cm
\noindent
If $A$ is a closed Hermitian operator, there exists a $1-1$ correspondence between
closed symmetric extensions $B$ of $A$, and partial isometries $V$, with initial space
${\cal H}_{I}(V) \subset {\cal K}_{+}$ and final space 
${\cal H}_{F}(V) \subset {\cal K}_{-}$. This correspondence is expressed by
$$
U_{B}=U_{A}+V,
$$
or by
$$
D(B)= \biggl \{ \varphi+\psi+V \psi: \varphi \in D(A),
\psi \in {\cal H}_{I}(V) \biggr \},
$$
where $B$ equals the restriction of the adjoint $A^{*}$ to the domain $D(B)$.
The operator $B$ is self-adjoint if and only if
$$
{\cal H}_{I}(V)={\cal K}_{+}, \; {\cal H}_{F}(V)={\cal K}_{-}.
$$
In particular, the operator $A$ has self-adjoint extensions if and only if 
$d_{+}=d_{-}$ and, in that case, if $d_{+} < \infty$, the set of self-adjoint
extensions is a $d_{+}^{2}$-dimensional real topological manifold, in
the topology of norm-resolvent convergence.} 
that provides the esplicit expression of all possible 
domains of the closed symmetric extensions for a 
closed symmetric operator, by using partial isometries between the 
deficiency spaces $K_{+}({\cal L})$ and $K_{-}({\cal L})$. 
Of course ${\cal L}$ is closable, thus we can make use of von Neumann's Theorem.

\noindent
Let us define the following unitary operator between the deficiency spaces 
that acts like a complex conjugation:
$$
{\cal U}:\xi\in K_{+}({\cal L})\longrightarrow{\cal U}\xi=\bar\xi\in K_{-}({\cal L}).
\eqno (5.16)
$$
Certainly ${\cal U}$ is an isometry, hence it is bijective.

\noindent From von Neumann's Theorem we know that the self-adjoint extension related 
to the unitary operator (5.16) has got the following domain:
$$
D(\hat{\cal L})=\{\eta+\xi+{\cal U}\xi\;|\;\eta\in D(\bar{\cal L}),\;\xi\in K_{+}({\cal L})\}. 
\eqno (5.17)
$$
It is straightforward that the function in $D(\hat{\cal L})$ satifies the 
relation (5.15) for same $\alpha$ and $\beta$.

\noindent Eventually, if we show that
$$
{\cal W}(x;\bar\phi,\psi)=0\hskip1cm\forall\phi,\psi\in D({\cal L}^{*})
\eqno (5.18)
$$
at the LP end-points, under our hypothesis of LP at both ends of $I$, 
from (5.12) we get the symmetry of the operator 
${\cal L}^{*}$, and because we must have $D({\cal L}^{*})\subseteq D({\cal L}^{**})$ 
but at the same time is easily seen that 
${\cal L}={\cal L}^{**}$ and $D({\cal L})\equiv D({\cal L}^{**})$, 
we must have $D({\cal L})=D({\cal L}^{*})$ and there follows 
that $\cal L$ is self-adjoint, thus its closure $\bar{\cal L}$ is symmetric from 
propositions (c) and (b) and has got 
deficiency indeces $n_{+}(\bar{\cal L})=n_{-}(\bar{\cal L})=0$.

\noindent In order to show that ${\cal W}(x;\bar\phi,\psi)=0$ at the LP extremes, 
suppose $b$ in LP. First of all we see that 
${\cal W}(b;\bar\phi,\psi)=0$ for all $\phi,\psi\in D(\hat{\cal L})$ because from 
(5.15) we have ${\cal W}(a;\bar\phi,\psi)=0$ 
and from the self-adjoint behaviour of $\hat {\cal L}$ we must have 
${\cal W}(b;\bar\phi,\psi)=0$ by relation (5.12). 
Now, if some $\eta_0\in C^{\infty}_{0}(I)$ is chosen in such a way that it 
equals zero on $[c,b)$ for some $c\in I$, and for which 
(5.15) is not verified, such a function must belong to 
$D({\cal L}^{*})\backslash D(\hat{\cal L})$. From the fact that 
${\cal L}$ has deficiency indices equal to one, there must be 
${\rm dim}(D({\cal L}^{*})-D(\hat{\cal L}))=1$ and therefore every 
function $\phi\in D({\cal L}^{*})$ can be written in the form 
$$
\phi=\phi_{0}+\lambda\eta_{0}\hskip1cm \lambda\in{\bf C},\;\;\;\phi_{0}\in D(\hat{\cal L}).
\eqno (5.19)
$$
Since $\eta_{0}(x)=\eta'_{0}(x)=0$ on $[c,b)$, then 
$\displaystyle\lim_{x\rightarrow b}{\cal W}(x;\bar\phi,\psi)
=\lim_{x\rightarrow b}{\cal W}(x;\bar\phi_{0},\bar\psi_{0})=0$ for all 
$\phi,\psi\in D({\cal L}^{*})$ of the form (5.19).

\noindent This completely shows that (5.18) holds, and the desired proof is completed.
\vskip 0.3cm
\leftline {\bf 6. Examples of self-adjoint extensions}
\vskip 0.3cm
\noindent
Here we want to show that indeed, the operator $\displaystyle-{ d^{2}\over dx^{2}}$ has got more that one 
self-adjoint extension, and these correspond to the Dirichlet and Neumann conditions at the origin. 
We can proceed in the following way: \par
\vskip0.2cm

1)\hskip0.5cm First we consider a particular class of domains -i.e.- 
$D_{\{ \epsilon,\mu\}}$, that let our operator 
be closed and symmetric and from this, using some arguments contained in [2], we are ensuring the 
existence of self-adjoint extensions for such closed and symmetric restrictions. \par
\vskip0.2cm

 2)\hskip0.5cm We use the von Neumann's theorem [2,3,14] to obtain explicitly all 
domains of closed and symmetric extensions 
-i.e- $D_{\{\epsilon,\mu\}}(c)$ (where $c$ runs over $[0,2\pi[$).
\vskip0.2cm

3)\hskip0.5cm We derive the form of the domains of the adjoint -i.e.- $D^{*}_{\{\epsilon,\mu\}}(c)$.
\vskip0.2cm

 4)\hskip0.5cm We use the Self-Adjointness condition $D_{\{\epsilon,\mu\}}(c)=D^{*}_{\{\epsilon,\mu\}}(c)$ to find 
which of the $D_{\{\epsilon,\mu\}}(c)$ domains is of self-adjointness.
\vskip0.2cm

Let us define the following two-parameter domains of symmetry for $\displaystyle-{ d^{2}\over dx^{2}}$:

$$
D_{\{\epsilon,\mu\}}=\{\phi\in C_{0}^{\infty}({\bf R}^{+}):\phi(x)=0 \; \forall x\in(0,\epsilon),
\; \phi(\epsilon)=\mu, \; \mu\in{\bf C}\}.
\eqno (6.1)
$$
It is easy to see that such domains are closed and on them our operator is symmetric, and from the fact that $n_{+}=n_{-}=1$, 
there exist self-adjoint extensions for each fixed admissible pair $(\epsilon,\mu)$.

\noindent In order to use von Neumann's Theorem, we need the expression of the deficiency spaces, and we easily find that
$$
{\cal K}_{+}=\{\phi\in{ L}^{2}({\bf R}^{+})\;|\;\phi=c_{+}\,e^{{i-1\over \sqrt{2}}x}\;,\hskip0.2cmc_{+}\in{\bf C}\}
\eqno (6.2)
$$
$$
{\cal K}_{-}=\{\phi\in{L}^{2}({\bf R}^{+})\;|\;\phi=c_{-}\,e^{-{i+1\over \sqrt{2}}x}\;,
\hskip0.2cmc_{-}\in{\bf C}\}\hskip0.2cm.
\eqno (6.3)
$$

\noindent We see that such spaces are one-dimensional linear spaces and, from von Neumann's Theorem we know that 
all possible symmetric extensions for each of $D_{\{\epsilon,\mu\}}$, are in bijection with the isometries 
between the deficiency spaces. From the fact that the deficiency spaces are one-dimensional, the isometries required can 
only be phase factors of the form $e^{i\theta(x,c_{+},c_{-})}$ and therein, 
following the statement of the Theorem, we must have
$$ 
e^{i\,\theta(x,c_{+},c_{-})}e^{{i-1\over \sqrt{2}}x}={c_{-}\over c_{+}}\,e^{-{i+1\over \sqrt{2}}x},
\eqno (6.4)
$$
and since it follows that $\biggl|{\displaystyle {c_{-}\over c_{+}}}\biggr|=1$, we set 
${\displaystyle {c_{-}\over c_{+}}}=e^{ic}$ with $c\in[0,2\pi[$, and this shows that $\theta (x,c)=-\sqrt 2 x+c$.

\noindent From the von Neumann's criterion [2,3,14] we can give the explicit form to the domains of symmetric 
extensions that we will call $D_{\{\epsilon,\mu\}}(c)$:
$$
D_{\{\epsilon,\mu\}}(c)=\bigl\{\psi\in{ L}^{2}({\bf R}^{+})\;|\;\psi=\phi+z\,(e^{{i-1\over \sqrt{2}}x}
+e^{ic}e^{-{i+1\over \sqrt{2}}x}) \bigr\},
\eqno (6.5)
$$
where $\phi \in D_{\{\epsilon,\mu\}}, z \in {\bf C}$.
Taking into account the symmetry relation 
$$
\left \langle \xi,\displaystyle{d^{2}\over dx^{2}}\psi \right \rangle
= \left \langle \displaystyle{d^{2}\over dx^{2}}\xi,\psi \right \rangle,
$$
where $\psi\in D_{\{\eta,\mu\}}(c)$ and 
$\xi\in D^{*}_{\{\eta,\mu\}}(c)$, 
we obtain the following equation:
$$
\xi^{'}(0)\,(e^{ic}+1)+\xi(0)\biggl({i+1\over\sqrt 2}e^{ic}-{i-1\over \sqrt 2}\biggr)=0\hskip0.2cm
\eqno (6.6)
$$
that defines the two following kinds of adjoint domains:
$$
D^{*}_{1}(c)=\biggl\{\xi\in{L}^{2}({\bf R}^{+})\;\biggl|\;{\xi(0)\over \xi^{'}(0)}=\,-\,{\sqrt{2}
\,(e^{ic}+1)\over 1+e^{ic}+i(e^{ic}-1)} \biggr\}
\eqno (6.7)
$$
with $c \not = {\pi \over 2}$, and
$$
D^{*}_{2}(c)=\biggl\{\xi\in{ L}^{2}({\bf R}^{+})\;\biggl|\;{\xi^{'}(0)\over \xi(0)}=\,-\,
\biggl({\sqrt{2}\,(e^{ic}+1)\over 1+e^{ic}+i(e^{ic}-1)}\biggr)^{-1}\biggr\},
\eqno (6.8)
$$
with $c \not=\pi$, in which we have ruled out 
${\pi \over 2}$ and $\pi$ values that lead to singular ratios
$\displaystyle{\xi(0)\over \xi^{'}(0)}$ and $\displaystyle{\xi'(0)\over \xi(0)}$, respectively.
First of all, it is interesting to note that (6.7) and (6.8) are independent of the $(\epsilon,\mu)$ pair. In this way 
we can certainly say that (6.7) and (6.8) cover {\it all possible domains for the adjoints of the closed and symmetric 
extensions for any of the possible closed and symmetric realizations of $-\displaystyle {d^{2}\over dx^{2}}$ 
over the real half-line}.

\noindent Now, by using the self-adjointness relations $D_{\{\epsilon,\mu\}}(c)=D_{1}^{*}(c)$ and 
$D_{\{\epsilon,\mu\}}(c)=D_{2}^{*}(c)$ we easily get the following self-adjointness domains:
$$ 
D^{*}_1({\pi})=D_{1}(\pi)= \{\xi\in{L}^{2}({\bf R})\;|\;\xi(0)=0\},
\eqno (6.9)
$$
$$ 
D^{*}_{2}\left({\pi \over 2}\right)=D_{2}\left({\pi \over 2}\right)
= \{\xi\in{L}^{2}({\bf R})\;|\;\xi^{'}(0)=0\},
\eqno (6.10)
$$
that correspond to the Dirichlet and Neumann condition at the origin.

\noindent The last thing that we want to note is that the sets (6.9) and (6.10) are both closed and open.

\noindent  For example, by using the following sequence in $D_{1}(\pi)\cap D_{2}\left({\pi\over 2}\right)$:
$$ 
f_{n}(x)=\cases{\;{\root 2 \of {{x}^{3}}}\hskip1cm x\in\displaystyle\biggl[0,{1\over n}
\biggr[\cr\displaystyle\;{1\over \root 3 \of x}\hskip1cm x\in\biggl]{1\over n},a
\biggr[\cr\;0\hskip1.4cm x\in[a,\infty[\cr}\hskip0.8cm a\in(1,\infty)
\eqno (6.11)
$$
which converges in $L^{2}$ but not in the intersection of $D_{1}(\pi)$ and
$D_{2}\left({\pi \over 2}\right)$, we realize that (6.9) and (6.10) are open sets.
On the other hand, if we choose the following:
$$ 
g_{n}(x)=\cases{\;{\displaystyle 1\over n}-\biggl(x-\displaystyle{1\over n}\biggr)^{2}
\hskip1cm x\in[0,a[\cr\;0\hskip2.9cm x\in[a,\infty)\cr}\hskip0.8cm a\in{\bf R}^{+}
\eqno (6.12)
$$
we see that $g_{n}\in \biggl( L^{2}({\bf R}^{+})-\bigl(D_{1}(\pi)\cup D_{2}\left({\pi \over 2})\right)\biggl)$, 
hence it belongs to the complement of each $D_{1}(\pi)$ and 
$D_{2}\left({\pi \over 2}\right)$ for every $n\in {\bf N}$ but 
its limit belongs $D_{1}(\pi)\cap D_{2}\left({\pi \over 2}\right)$. 
This shows that the complement of (6.9) and (6.10) is 
an open set and therefore (6.9) and (6.10) must be closed sets. 

\noindent The fact that they are closed sets also results from von Neumann's Theorem. In this way, the 
sequence (6.12) confirms the validity of such a Theorem.
Eventually, we have obtained that the operator $-\displaystyle {d^{2}\over dx^{2}}$ is 
Self-Adjoint only on domains (6.13) and (6.14), which are simultaneously closed and open.
\vskip 0.3cm
\leftline {\bf 7. Concluding remarks}
\vskip 0.3cm
\noindent
In the first part of our paper we have derived two peculiar aspects of the general Sturm-Liouville operators.
First over all, starting from very general and fundamental theorems, we have shown which are the weakest 
assumptions on the coefficients in order to obtain solutions of the eigenvalue 
problem that are sufficiently regular, i.e. 
-absolutelly continuous- to be used whenever needed. A second remarkable aspect is essentially seen 
in the possibility of considering such regular solutions embedded in a hightly non-regular space such as (4.3).
It is from this latter aspect that our proof of Weyl's theorem takes the moves. Other proofs of this theorem
can be found in [15,16], under the natural assumption of differential operators
acting on a suitable Hilbert space. As far as this last pair of references are concerned, we want to 
mention some further functional-analytical 
methods which are up to date with the current developments of this subject. An example is the 
{\it maximal operator} and {\it minimal operator}
related to a differential expression like (2.1) or (4.2). In [15], the domain of definition  
for the maximal operator related to a $n^{th}$-order
differential expression $\tau$ on a real line segment is
$$
D_{M}=\{ f\in{\cal L}^{2}(a,b)\,:\,f^{(0)},f^{(1)}, ... ,f^{(n-1)}\in AC(a,b); 
\;\tau f\in{\cal L}^{2}(a,b) \} , 
$$
and it is shown that such a domain is densely defined and closed in the Hilbert space.
Here we also define the domain for the minimal operator in the form
$$
D_{m}=\{f\in {\cal L}^{2}(a,b)\,:\,f\in C_{0}^{2}(a,b)\}.
$$
Upon focusing on our differential operator (4.2), it is evident that the minimal operator 
occurs instead under the hypothesis of Theorem 4.1.
With the language of our paper and by means of Theorem 4.2, 
the maximal set $D_{M}$ is basically 
$$
D_{M}=\{f\in W^{2,1}_{\rm loc}([a,b])\,:\,f^{''}\in {\cal L}^{2}(a,b)\},
$$
hence we derive the Hilbert-space nature of $D_{M}$ by relying only upon the square 
summability of highest derivatives in it.
In our proof we decided to use the {\it Von Neumann's extension theorem} in order 
to reach all possible self-adjoint extensions 
for the specific differential operator under consideration. 
We suggest reading [15] for a the general 
theory about $n^{th}$-order differential
operators's self-adjoint extensions in terms of boundary conditions of which 
our (5.18) represents a specific case. We remark the 
fact that, with the language used here, the expression (5.18) cannot be untied from the functional space (5.10).
Following the conceptual behaviour of Von Neumann's theorem, 
we must bring to the attention of the reader another functional method
that makes use of a tool called {\it boundary triples}. Given any Hilbert space and a symmetric operator $T$,
it is always useful to define a sequilinear map $\Gamma$
$$
\Gamma_{T}(\phi,\psi)= \langle T^{*}\phi,\psi \rangle 
- \langle \phi,T^{*}\psi \rangle
$$
for $\phi,\psi\in D(T^{*})$ in order to find the closure of $T$. We also refer to [16] 
for a recent redefinition of essential self-adjointness
in terms of the map $\Gamma$. By the way, a boundary triple is a triplet $(h,\rho_{1},\rho_{2})$ 
where $h$ is a suitable Hilbert space and $\rho_{1},\rho_{2}$ 
are $D(T^{*})\longrightarrow h$ maps that satisfy the rule 
$$
a \Gamma_{T^{*}}(\phi,\psi)
= \langle \rho_{1}(\phi),\rho_{1}(\psi)\rangle  
- \langle \rho_{2}(\phi),\rho_{2}(\psi) \rangle
$$
for some complex $a$ constant. In terms of boundary triples, one finds that $\Gamma(\phi,\psi)$ 
vanishes identically when we restrict the inputs $\phi,\psi$ to the domain of
some self-adjoint extension. One can therefore say that the domain of any self-adjoint 
extension of a symmetric operator $T$ is of the form
$$	
D(T_{\cal U})=\{\phi\in T^{*}\,:\,\rho_{2}(\phi)={\cal U}\rho_{1}(\phi)\}
$$
where $\cal U$ is a unitary operator. In terms of boundary triples, one can 
obtain a more general statement and expression 
for the self-adjoint extensions of a symmetric operator. The role played by 
the sesquilinear map $\Gamma$ is the same as the role played by the domain (5.10).
Such a theory is described in detail in [16].

To sum up, we have indeed shown that the operator (4.2) can clearly act over a sort 
of spaces like (5.1), and that the Hilbert-space nature of such sets, resides only in the square summability 
of second derivatives of their functions by virtue of Theorem 4.2.
In this fashion, whenever needed, one can look for weaker solutions of the eigenvalue equation.

Similar techniques have been applied, over the years, to a wide range of topics. For example,
the work in [17] studied essential self-ajointness in $1$-loop quantum cosmology, 
the work in [18] has provided enlightening examples of boundary conditions 
for self-adjoint extensions of linear operators, whereas the
work in [19] has suggested that a profound link might exist between the formalism for
asymptotically flat space-times and the limit-point condition for singular Sturm-Liouville
problems in ordinary quantum mechanics. Last, but not least, the parameter 
$\lambda_{nl} \equiv l+{(n-2)\over 2}$ in Eq. (1.2) is neatly related to the parameter $L$
used in large-$N$ quantum mechanics [20], i.e.
$$
\lambda_{nl}={L \over 2}-1.
$$
Moreover, since the Schr\"{o}dinger stationary states are even functions of $\lambda_{nl}$, this 
suggests exploiting the complex-$\lambda_{nl}$ plane in the analysis of scattering problems [21].
If $n$ is kept arbitrary, this means complexifying a linear combination of $l$ and $n$ [5],
including the particular case where $l$ remains real while the dimension $n$ is complexified.

Thus, there is encouraging evidence that Sobolev-space methods 
and yet other concepts of functional and complex analysis may provide the appropriate tool
for investigating classical and quantum physics as well as correspondences among such frameworks.
\vskip 0.3cm
\noindent
{\bf Acknowledgments}.
V F B is grateful to G E for support and useful advice.
G E is grateful to Dipartimento di Fisica ``Ettore Pancini'' for hospitality and support.
\vskip 0.3cm
\noindent
ORCID of G Esposito: 0000-0001-5930-8366
\vskip 0.3cm
\leftline {\bf G-regularity}
\vskip 0.3cm
\noindent
For the definition of G-regularity we need to endow the ${\bf R}^{n}$ space with the norm
$$
||y||=\sum_{i=1}^{n}|y_{i}|.
\eqno (A1)
$$
Here we make use of the topology induced by the {\it uniform convergence} in order 
to give to the $C(I,{\bf R}^{n})$ vector 
space a Banach structure. Let us call it ${\cal K}$.

\noindent We will say that the sequence $u_{n}\in {\cal K}\longrightarrow u\in {\cal K}$ 
if and only if, for every $\epsilon\in{\bf R}^{+}$, 
there exists some $\nu\in{\bf N}$ such that, for all $n\geq \nu$ 
the following majorization is verified:
$$
{\rm sup}_{I}|u_{n}(x)-u(x)|<\epsilon.
\eqno (A2)
$$
Since $\cal K$ is also a metric space, we can use in (A2) the following 
notation for the distance between $u_{n}$ and $u$:
$$
d_{n}={\rm sup}_{I}|u_{n}(x)-u(x)|.
\eqno (A3)
$$
At this stage we have only to define the following particular sequence on the $I$ segment:
$$
x^{n}=x-d_{n}{\rm sign}(x)
\eqno (A4)
$$
and we note that $x^{n}\longrightarrow x$ on $I$ only if $u_{n}\longrightarrow u$ on $\cal K$.

\noindent We are now in a position to define the concept of G-regularity,
while we refer to Definition 2 for the G-integrability.
\vskip 0.3cm
\noindent
{\bf Definition 6.}\hskip0.5cm Let $f$ be G-integrable on the interval $I$ whose middle point is $x_{0}$, 
and let $g\in G$ be defined in equation (2.3). 
Let the sequence $u_{n}(x)=\int_{x_{0}}^{x}g_{n}(s)ds\in G$ be such that $u_{n}$ tends uniformly to 
$u(x)=\int_{x_{0}}^{x}g(s)ds$ on $I$. Let $x^{n}$ be defined by (A4). 

\noindent If now every such sequence of functions verifies the condition
$$
f\bigl(x^{n},u_{n}(x^{n})\bigr)\rightarrow f\bigl(x,u(x)\bigr),
\eqno(A5)
$$
then we say that $f$ is {\it G-regular} on $I$.
\vskip0.5cm
For more insights on the G-regularity property of functions, we refer the reader to [12].
\vskip 0.3cm
\leftline {\bf References}
\vskip 0.3cm
\noindent
\item {[1]}
Weyl H 1910 {\it Mathematische Annalen} {\bf 68} 220

\item {[2]}
Reed M and Simon B 1975 {\it Methods of Modern Mathematical
Physics. II. Fourier Analysis and Self-Adjointness}
(New York: Academic)

\item {[3]}
Simon B 2015 {\it A Comprehensive Course in Analysis, Part 4} 
(Providence: American Mathematical Society)

\item {[4]}
Lieb E H and Loss M 1997 {\it Analysis} (Providence: American Mathematical Society)

\item {[5]}
Esposito G 1998 {\it Found. Phys. Lett.} {\bf 11} 535

\item {[6]}
Coddington E A and Levinson N 1955 {\it Theory of Ordinary Differential Equations} (New York: McGraw-Hill)

\item {[7]}
Yosida K 1960 {\it Lectures on Differential and Integral Equations}
(New York: Interscience Publishers)

\item {[8]}
Kurss H 1967 {\it Proceedings of the American Mathematical Society} {\bf 18} 445

\item {[9]}
Brinck I 1959 {\it Mathematica Scandinavica} {\bf 7} 219

\item {[10]}
Everitt W N Knowles I W Read T T 1986 {\it Proceedings of the Royal Society of Edinburgh} {\bf 103A} 215

\item {[11]}
Aquaro G 1955 {\it Bollettino Unione Matematica Italiana} {\bf 10} 208

\item {[12]}
Persson J 1975 {\it Journal of Mathematical Analysis and Applications} {\bf 49} 496

\item {[13]}
Brezis H 1986 {\it Functional Analysis} (New York: Springer)

\item {[14]}
Dell'Antonio G 2011 {\it Mathematical Aspects of Quantum Mechanics. I} (Naples: Bibliopolis)

\item {[15]}
We\-id\-ma\-nn J 1987 {\it Sp\-ec\-tr\-al Th\-eo\-ry o\-f 
Or\-di\-na\-ry Di\-ff\-er\-en\-ti\-al Op\-er\-at\-or\-s}
(Be\-rl\-in: Sp\-ri\-ng\-er)

\item {[16]}
de Oliveira C R 2009 {\it Intermediate Spectral Theory and Quantum Dynamics}
(Boston: Birkh\"{a}user)

\item {[17]}
Esposito G, Morales-T\'ecotl H A and Pimentel L O 1996 
{\it Class. Quantum Grav.} {\bf 13} 957; erratum {\it ibid.} {\bf 17} 3091

\item {[18]}
Asorey M, Ibort A and Marmo G 2005 {\it Int. J. Mod. Phys.} {\bf A20} 1001

\item {[19]}
Esposito G and Alessio F 2018 {\it Gen. Relativ. Gravit.} {\bf 50} 141

\item {[20]}
Chatterjee A 1990 {\it Phys. Rep.} {\bf 186} 249

\item {[21]}
De Alfaro V and Regge T 1965 {\it Potential Scattering} (Amsterdam: North Holland)

\bye